\documentclass[
    ,final            
  ]
  {aipproc}

\layoutstyle{6x9}

\def\beq{\vspace{-0.3em}\color{magenta}\begin{eqnarray*}}
\def\eeq{\end{eqnarray*}\color{blue}\vspace{-0.3em}}
\def\bea{\begin{eqnarray*}}
\def\eea{\end{eqnarray*}}

\def\lQ{\Lambda_{\rm QCD}}

\def\als{\alpha_{\rm s}}

\def\siml{{\ \lower-1.2pt\vbox{\hbox{\rlap{$<$}\lower6pt\vbox{\hbox{$\sim$}}}}\ }}     
\def\simg{{\
    \lower-1.2pt\vbox{\hbox{\rlap{$>$}\lower6pt\vbox{\hbox{$\sim$}}}}\ }}

\newcommand{\be}{\begin{equation}}
\newcommand{\ee}{\end{equation}}

\begin{document}

\title{QCD Effective Field Theories for Heavy Quarkonium}


\author{Nora Brambilla}{
  address={Dipartimento di Fisica dell'Universit\`a di Milano and INFN}
}

\begin{abstract}
QCD nonrelativistic effective field theories (NREFT)
are  the modern and most suitable frame to describe  heavy
quarkonium properties. Here  I summarize
few  relevant concepts and some of the interesting physical applications
(spectrum, decays, production)  of NREFT.
\end{abstract}

\maketitle

\vspace{-15mm}
\section{Introduction}

The study of heavy quark-antiquark bound states touches upon several
important physics areas, within and beyond the Standard Model of
Particle Physics.  These
multi-scale systems probe all the energy regimes of QCD, from the hard
region, where an expansion in the coupling constant is legitimate, to
the low energy region, where nonperturbative effects dominate. 
Heavy quark-antiquark states  are  thus an ideal and to some extent unique 
laboratory where our understanding of nonperturbative QCD 
and its interplay with perturbative QCD may be tested in a controlled framework.
This has been so historically, since quarkonium has been deeply related to 
the development of QCD, and it is so  even more today  for the two following  main reasons.
The first reason  is that 
in the last few years a wealth of new experimental results has become available.
The diversity, quantity and accuracy of the data currently being collected is impressive 
and includes:
data on quarkonium formation from BES at BEPC, E835 at Fermilab,
KEDR (upgraded) at VEPP-4M, and CLEO-III, CLEO-c; 
clean samples of charmonia produced in B-decays, in photon-photon
fusion and in initial state radiation, from the B-meson factory
experiments, BaBar at SLAC and Belle at KEK, including the unexpected
observation of large amounts  of associated $(c\overline{c})(c\overline{c})$ production;
the CDF and D0 experiments at Fermilab measuring heavy quarkonia
production from gluon-gluon fusion in $p\bar{p}$ annihilations at
2~TeV; ZEUS and H1, at DESY, studying charmonia production in photon-gluon
fusion;
PHENIX and STAR, at RHIC, and NA60, at CERN, studying charmonia
production, and suppression, in heavy-ion collisions.
In the near future, even larger data samples are expected from the
BES-III upgraded experiment, while the B factories and the
Fermilab Tevatron  will continue to supply
valuable data for several years.  Later on, new facilities will become
operational (LHC at CERN, Panda at GSI, much higher luminosity B
factories at KEK, a Linear Collider, etc.) offering fantastic
challenges and opportunities in this field.  
A  comprehensive review of the experimental and theoretical 
status of heavy quarkonium physics may be found in the Cern Yellow Report prepared by 
the Quarkonium Working Group (http://www.qwg.to.infn.it)
\cite{qwg}. See also the experimental review talks \cite{exptalks} at this conference.

The second reason is the remarkable theoretical progress of the last few years.
Effective field theories (EFT), such as Nonrelativistic QCD (NRQCD)\cite{nrqcd1,nrqcd2},
provided new tools and definite predictions concerning, for instance,
heavy quarkonium production and decays.  New effective field theories
for heavy quarkonium, as potential NRQCD (pNRQCD)\cite{pnrqcd1,pnrqcd2} and velocity NRQCD
(vNRQCD)\cite{vnrqcd}, have been recently developed and are producing a wealth of
new results.  The lattice implementation of such effective theories
has been partially carried out and many more results with drastically
reduced systematic uncertainties are expected in the near future. 
An extensive  review of the latest development in nonrelativistic EFTs can be found 
in \cite{reveft}.

Therefore, on one hand the progress in our understanding of NREFTs makes it 
possible to move beyond phenomenological models and to  provide a systematic description 
from QCD of all aspects of heavy-quarkonium physics. On the other hand, the recent 
progress in the measurement of several heavy-quarkonium observables makes it meaningful 
to address the problem of their precise theoretical determination.
In this situation heavy quarkonium becomes a very special and relevant system to advance 
our understanding of strong interaction and our control of some parameters of the Standard
Model. 

In the following we present the main conceptual ideas and simplifications 
underlying the EFT framework, a brief summary of the main ingredients 
of NRQCD and  pNRQCD and list several applications to the  phenomenology of quarkonium,
referring to the original publications for the details.
\vspace{-3mm}
\section{Nonrelativistic Effective Field Theories}
\vspace{-1mm}
\subsection{Nonrelativistic bound states}

\begin{figure}
\makebox[0truecm]{\phantom b}
\put(-140,0){\includegraphics[height=.07\textheight]{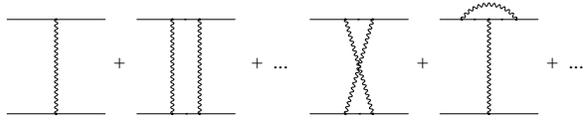}}
  \caption{QED series of diagram defining the $e^+ e^-$ interaction.
 The resummation of the series of all ladder photon diagrams (of the type of the two 
 first diagrams):  $\alpha\left(1+{\alpha\over v}  + \dots \right)$ (with $\alpha \sim v$)
 gives the (leading order) Coulomb bound state energy pole.}
\vspace{-400mm}
\label{fig:uno}
\end{figure}

Nonrelativistic bound states are characterized by a small  relative 
velocity $v$ (in the centre-of-mass frame) of the particles inside the bound system.
The bound state dynamics has quite distinctive features with respect to the scattering
 case.
If one considers the simpler example of a QED nonrelativistic bound system
like  positronium (or hydrogen), then $v \simeq \alpha \ll 1$. The 
fine structure constant $\alpha$ being  small in QED, 
physical observables  may be evaluated in perturbation theory.
However, the bound state  is nonperturbative and in this case
one needs to resum infinite sets of Feynman diagrams. 
Due to the fact that the bound state lives close to threshold, i. e. $v \simeq \alpha$,
the bound state pole (at leading order) is obtained by the 
resummation of all the Coulomb ladder photon contributions,  $ \simeq ({\alpha\over v})^n$,
in the Feynman diagram series 
(cf. Fig.\ref{fig:uno}).  As a consequence by solving  the Schr\"odinger equation             
$({p^2\over 2m } + V)\phi =E \phi $, with $V$ a  Coulomb potential,
 one  generates the dynamical scales of the  momentum transfer
$p \simeq m \alpha$ and of the bound state kinetic energy $ E= {p^2\over m}\simeq m \alpha^2$  
(for some review see \cite{nreftn}).   Such scales manifest themselves  
in the scalings of positronium  radial  
($\sim  m \alpha^2  $), fine and hyperfine splittings ($\sim m \alpha^4 \ $).
This is different from a pure scattering calculation
where no scales involving $\alpha$ are generated.
\vspace{-7mm}

\subsection{Quarkonium scales}

Heavy quarkonium  provides a  non-relativistic system 
potentially very similar to a QED bound state,
with the difference that the  low-energy scales 
will be sensitive to $\Lambda_{\rm QCD}$.
Therefore the description of hadrons containing 
two heavy quarks is a rather challenging problem,  
which adds to the complications of the bound state in field theory 
those coming from a nonperturbative QCD low-energy dynamics.
A  simplification is provided  by  the nonrelativistic nature of heavy quarkonium,
suggested by the large mass of the heavy quarks
and  manifest in the spectrum pattern.
Quarkonium is thus characterized
 by three energy scales, hierarchically ordered by the quark velocity 
$v \ll 1$: the mass $m$ (hard scale), the momentum transfer $mv$ (soft scale), 
which is proportional to the inverse of the typical size of the system $r$,
 and  the binding energy $mv^2$ (ultrasoft scale), which is proportional to the 
inverse of the typical 
time of the system. In bottomonium $v^2 \sim 0.1$, in charmonium $v^2 \sim 0.3$.
In perturbation theory $v\sim \als$. Feynman diagrams will get contributions 
from all momentum regions associated with the scales.
Since these momentum regions depend on $\als$, each Feynman diagram contributes 
to a given observable with a series in $\als$ and a non trivial counting.
For energy scales close to $\lQ$, the scale at which 
nonperturbative effects become dominant, perturbation theory breaks down 
and one has to rely on nonperturbative methods.
 Regardless of this, the non-relativistic hierarchy 
$ m \gg mv \gg mv^2 $ will persist  also below  the  $\lQ$  threshold.

The wide span of energy scales involved makes also a lattice calculation in full QCD 
extremely challenging since one needs a space-time grid that is large compared 
to the largest length of the problem, $1/mv^2$, and a lattice spacing that 
is small compared to the smallest one, $1/m$.  To simulate, for instance, a $b\bar{b}$ 
state where $m/mv^2\sim 10$, one needs lattices as large as $100^4$, which are extremely 
time computing demanding.

We may, however, take advantage of the existence of a hierarchy of scales  
by substituting QCD with simpler but equivalent EFTs.
A hierarchy of EFTs may be constructed by systematically integrating out 
modes associated to  energy scales not relevant for the quarkonium system.
Such integration  is made  in a matching procedure that 
enforces the complete equivalence between QCD and the EFT at a given 
order of the expansion in $v$
and achieves a  factorization 
between the high energy and the low energy contributions.
\vspace{-1mm}
\subsection{What is an EFT?}
Let $H$ be a system described by a fundamental Lagrangian $\cal L$. 
Suppose  that  $H$   has two characteristic physical scales:  $\Lambda \gg
\lambda$. The EFT Lagrangian, ${\cal L}_{\rm EFT}$, suitable to describe $H$ 
at scales lower than $\Lambda$  is then defined by: 
a cut off   $\Lambda \gg \mu \gg \lambda$;
some  degrees of freedom  that exist at scales lower than $\mu$.  
${\cal L}_{\rm EFT}$  involves   all the operators  $O_n$ 
that may be built from the effective degrees  
of freedom and are consistent with the  symmetries of $ {\cal L}$:
\begin{equation}
 {\cal L}_{\rm EFT}  = \sum_n  c_n(\Lambda, \mu)  {O_n(\mu,\lambda)
  \over \Lambda^n}
\label{left}
\end{equation}
\begin{itemize}
\item{}{Once the scale $\mu$ has been run down to $\lambda$, 
$\langle  O_n  \rangle \sim  \lambda^n$, so that the EFT is
  organized as an expansion in the small parameter $\lambda/\Lambda$.}
\item{} {The EFT is  renormalizable order by order in
  $\lambda/\Lambda$}.
\item{} {The  matching coefficients  $c_n(\Lambda, \mu)$  encode the
  non-analytic behaviour in  $\Lambda$. They are calculated by imposing 
that   ${\cal L}_{\rm EFT}$  and   ${\cal L}$  describe the same
  physics at any finite order in the expansion:  this is called matching procedure.}
\item{}{If $ \Lambda \gg \lQ$  then the $c_n(\Lambda, \mu)$ may be calculated in
   perturbation theory.}
\end{itemize}
The EFT has a well defined power counting in a small parameter 
so that the expansion is systematic.
In QCD  the EFT approach  makes it possible to achieve a rigorous factorization between 
 the high-energy dynamics encoded into matching coefficients 
calculable in perturbation theory and the nonperturbative QCD dynamics 
encoded  into few well-defined nonperturbative operator matrix elements 
to be fitted on the data or calculated on the lattice or in QCD vacuum models.
Thus, several model independent QCD predictions become possible. 
EFTs have the additional
advantage that they are often close to popular phenomenological models to which experimental results
are usually compared to, the main difference being a good control on the
systematic errors.
\vspace{-2mm}
\section{NRQCD}
 Integrating out degrees of freedom 
of energy $m$, which for heavy quarks can be done perturbatively, leads to 
NRQCD \cite{nrqcd1,nrqcd2}.  In the language of the previous section
this is the EFT that follows from QCD when $\Lambda = m$.  
The  matching is  perturbative.
The Lagrangian is organized as an expansion in $v$  and 
  $\als(m)$:
\begin{equation}
 {\cal L}_{\rm NRQCD}  = \sum_n  c_n(m,\mu)  \times  O_n(\mu, mv,mv^2,  \lQ)/m^n .
\label{eftnrqcd}
\end{equation}
The Wilson coefficients 
$c_n$ are series in $\als $ and encode the ultraviolet physics that  
has been integrated out from QCD. The operators $O_n$ 
describe the low-energy dynamics and are counted in powers of $v$.
However since two scales, the soft and the ultrasoft, still remain dynamical,
NRQCD does  not have an unambiguous  power counting in $v$.
The operators bilinear in the fermion (or in the antifermion) fields are the same 
that can be  obtained from a standard Foldy-Wouthuysen transformation \cite{rev}
(however corrected with the matching coefficients). 
In this framework  the imaginary part of the NRQCD Hamiltonian, i.e. the 
imaginary part of the Wilson coefficients of the 4-fermion operators,
is responsible for  heavy quarkonium annihilations.
Being this a field theory, the heavy quarkonium Fock state is 
given by a series of terms in which the leading one is a  $Q\bar{Q}$ 
in a color singlet state and the first correction, suppressed in $v$,
comes from a $Q\bar{Q}$ in an octet state with a gluon.

\leftline{\bf Applications of NRQCD}
\noindent
{\bf Spectrum.}
The NRQCD Lagrangian is well suited for lattice calculations. The quark propagators 
are the nonrelativistic ones and since we have integrated out the scale of the mass, 
the lattice step used in the simulation may be a factor $1/v$ bigger. Lattice evaluation 
of heavy systems like bottomonium become thus feasible. The latest results for the spectra 
(quenched and unquenched) are given e.g. in \cite{latticenrqcd}. The radial splittings are accurate 
up to $O(\als v^2)$ while fine and hyperfine splittings are accurate only up to $O(\als)$,
due to the fact that only tree level matching coefficients have been used. A calculation 
of the NRQCD matching coefficients in the lattice regularization scheme is still missing
and would be relevant to improve the precision of the lattice data.\par
\noindent
{\bf Decays.}
NRQCD gives a factorization formula for heavy
quarkonium  inclusive decay widths into light hadrons and electromagnetic decays
 involving four-fermion matching coefficients and  matrix elements of four fermion operators \cite{nrqcd2}.
Singlet operator expectation values may be easily related to the square 
of the quarkonium wave functions (or derivatives of it) at the origin. 
Octet contributions remain as  nonperturbative matrix elements 
of operators evaluated over the quarkonium states.
In some situations the octet contributions may not be suppressed and become as 
  relevant as the  singlet contributions in the NRQCD power counting. In particular  
octet contributions 
may reabsorb the dependence on the infrared cut-off  of the Wilson coefficients, 
solving the problem that arised in the color singlet potential model.
Systematic
improvements are possible, either by calculating higher-order corrections 
in the coupling constant or by adding higher-order operators. If one goes 
on in the expansion in $v$, that seems to be necessary for charmonium,
the number of  involved nonperturbative matrix elements of octet operators
 increases in such a  way that limits the prediction power \cite{decrev}.
In the matching coefficients large contributions in the perturbative series 
coming from bubble-chain diagrams may need to be resummed \cite{chen}.
\par\noindent
{\bf Production.}
Before the advent of NRQCD, colour singlet production and colour singlet 
fragmentation underestimated the data on prompt quarkonium production at Fermilab 
by about an order of magnitude indicating that additional fragmentation contributions 
were missing \cite{prod}. The missing contribution is precisely the gluon fragmentation into 
colour-octet $ ^3S_1$ charm quark pairs. The probability to form a $J/\psi$ particle from a pointlike 
$c\bar{c}$  pair in a colour octet  $ ^3S_1$ state is given by a NRQCD 
 nonperturbative matrix element 
which is suppressed by $v^4$ with respect to  to the leading singlet term but 
 is enhanced by two powers of $\als$ 
 in the short distance matching coefficient
  for producing colour-octet quark pairs. When one introduces the leading 
colour-octet contributions, then the data of CDF can be reproduced. 
Still remains a puzzle the behaviour of the polarization at high $p_{\rm T}$ (cf. the production 
chapter in \cite{qwg}).

\vspace{-5mm}
\section{pNRQCD}
\vspace{-2mm}
 In NRQCD   the role of the potential and the
 quantum mechanical nature of the problem are not yet maximally
 exploited. A higher  degree of simplification may  be achieved
building  another EFT 
where only the ultrasoft  degrees of freedom remain dynamical.
pNRQCD \cite{pnrqcd1,pnrqcd2,reveft} is the EFT for heavy quarkonium that follows from NRQCD when 
$\Lambda = 1/r \sim m v$.
We may distinguish two situations:
1) weakly coupled pNRQCD when $mv  \gg \lQ$, 
where the matching from NRQCD to pNRQCD may be performed in
 perturbation theory
2) strongly coupled pNRQCD when $ mv \sim \lQ$, 
where the matching has to be nonperturbative.
Recalling that $r^{-1}
  \sim mv$, these two situations correspond  to systems with inverse
typical radius smaller than or  of the same order as  $\Lambda_{\rm QCD}$.

\vspace{-7mm}
\subsection{Weakly coupled pNRQCD}
The effective degrees of
freedom are: low energy $Q\bar{Q}$ states (that can be decomposed into  a singlet
and an octet field under colour transformations) 
with energy of order  $ \Lambda_{\rm QCD},mv^2$ and
    momentum   ${\bf p}$ of order $mv$,  
plus  ultrasoft gluons
with energy  and momentum of order
$\lQ,mv^2$. 
All the  gluon fields are multipole  expanded (i.e.  expanded  in
$r$). 
The Lagrangian is then given by terms of the type
\begin{equation}
{c_k(m, \mu) \over m^k}  \times  V_n(r
\mu^\prime, r\mu)   \times  O_n(\mu^\prime, mv^2, \lQ)\; r^n  .
\end{equation}
where the potential matching 
coefficients $V_n$  encode the non-analytic behaviour in $r$.
At  leading order in the multipole expansion,
the singlet sector of the Lagrangian gives rise to equations of motion of the 
Schr\"odinger type. 
We point out that:
each term in  the pNRQCD Lagrangian has a definite power counting
and there is a systematic 
procedure to calcolate corrections in $v$ to physical observables;
higher order perturbative (bound state) 
calculations in this framework become much simpler \cite{pnrqcdpert,US}.
In particular the EFT can be used for a very efficient resummation of 
the large logs (typically logs of the ratio of energy and momentum scales)
 using the renormalization group  (RG) adapted to the case of correlated scales \cite{rg};
retardation (or non-potential) effects 
start at the NLO in the multipole expansion and are systematically 
encoded inside pNRQCD.  They are typically related to nonperturbative effects
\cite{pnrqcd2,reveft};
Poincar\'e invariance is not lost, but shows up 
in some exact relations among the matching coefficients \cite{poincare0}.
We emphasize that inside the EFT framework the renormalon subtraction 
may be implemented systematically. The  renormalon subtraction
 allows us to obtain convergent perturbative series  and to 
unambiguously define power corrections\cite{reveft}. This  is one of the main 
reasons for the success of the several applications listed below
(for any details see the quoted references). 
\vskip 0.05truecm
\leftline{\bf Applications of weakly coupled pNRQCD}
\noindent
{\bf QCD Singlet Static potential.}
The singlet and octet potentials are well defined objects 
to be calculated in the perturbative matching. In \cite{US}
a determination
of the singlet potential at three loops leading log has been obtained  
and correspondingly  also a determination of $\alpha_V$ showing how 
this quantity starts to depend on the infrared behaviour  of the theory at three loops.
The perturbative calculation of the static potential at (almost) three loops and with 
the RG improvement has been compared to the lattice calculation of the potential and found 
in good agreement up to about 0.35 fm \cite{pertlat}.
\par\noindent
{\bf $b$ and $c$ masses.} 
\par\noindent
Heavy quarkonium is one of the most suitable system to extract a precise 
determination of the mass of the heavy quarks $b$ and $c$.
Perturbative determinations of the 
$\Upsilon(1S)$ and $J/\psi$ masses have been used to extract the $b$ and $c$
masses. The main uncertainty in these determinations
comes from nonperturbative contributions (local and nonlocal condensates \cite{pnrqcd2})
together with possible effects due to
subleading renormalons. Table \ref{tableMSmasses} shows some recent determinations.
For a discussion about the errors and the difference 
among  the given results see \cite{reveft}.
A recent analysis  performed by the QWG \cite{qwg}
and based on all the previous determinations existing in the 
literature indicates that the mass extraction from  heavy quarkonium 
involves an error of about 50 MeV both in the bottom 
($1\% $ error) and in the charm ($4 \% $ error) case.
\par\noindent
{\bf Perturbative quarkonium  spectrum.}\par \noindent
{\bf $B_c$ mass}. Table \ref{TabBc} shows some 
recent determinations of the $B_c$ mass in perturbation theory at NNLO 
accuracy compared with a  recent 
lattice study \cite{Allison:2004be} and the value of the CDF  experimental  $B_c$ mass. 
This would support 
the assumption that nonperturbative contributions to the quarkonium ground state are of the 
same magnitude as NNLO or even NNNLO corrections, which would be consistent with a 
$ mv^2 \simg\lQ$ power counting.
{\bf Hyperfine splittings.}
$c\bar{c}$, $b \bar{b}$, $B_c$ ground state hyperfine splittings have been recently calculated 
at NLL in \cite{Kniehl:2003ap}. The prediction for $\eta_b$ mass is
$M(\eta_b) = 9421 \pm 10\,{(\rm th)} \,{}^{+9}_{-8}\, (\delta\als)~{\rm MeV}$.
The logs resummation seems to be important.
If the experimental error in future measurements of $M(\eta_b)$
  will not exceed few Mev, the bottomonium hyperfine separation will become a 
  competitive source of $\alpha_s(M_Z)$ with an estimated accuracy  of 
  $\pm 0.003$.
{\bf Higher bottomonium resonances} have been investigated in the framework of
perturbative QCD most recently in \cite{Brambilla:2001fw,Brambilla:2001qk,Recksiegel:2002za}.
The surprising result of these studies is that some gross features of the
lowest part of the bottomonium spectrum, like the approximate equal
spacing of the radial levels, are reproduced by a perturbative calculation
that implements the leading-order renormalon cancellation. If this is coincidental or reflects the
(quasi-)Coulombic nature of the states will be decided by further
studies. A recent NLO calculation of the 1$P$ bottomonium fine splittings 
has been performed in \cite{Brambilla:2004wu}. 
It seems to indicate either the existence of large NLL/NNLO
corrections (as it happens in the hyperfine splittings of the 1$S$ levels) 
or sizeable nonperturbative corrections.
\par\noindent
{\bf Radiative transitions (M1,E1).}
A theory of M1 and E1 transitions in heavy quarkonium has been recently formulated
using pNRQCD \cite{vairoh}. This  may shed some light on recent CLEO results on radiative M1 transitions
in the $\eta_b$ search that have ruled out several models.
\par\noindent
{\bf Seminclusive radiative decays of $\Upsilon(1S)$.} 
In \cite{radup} the end-point region of the photon spectrum in semi-inclusive radiative decays of 
 heavy quarkonium has been discussed using Soft-Collinear Effective Theory and pNRQCD.
Including the octet contributions a good understanding of the experimental  data is obtained 
\par\noindent
{\bf Top-antitop production  near threshold at ILC.}
In \cite{ttbar} the total cross section for top quark pair production 
close to threshold in e+e- annihilation is investigated at NNLL in vNRQCD.
The summation of logarithms leads to a convergent expansion for the normalization of 
the cross section, and small residual scale dependence.
This makes precise extractions of the strong coupling, the top mass 
and the top width feasible.
\par\noindent
{\bf  Determinations of  $\als$.}
Heavy quarkonia leptonic and non-leptonic inclusive decays 
  rates may provide means to extract $\alpha_s$. The present 
  PDG determination of $\alpha_s$ from heavy quarkonium pulls down 
  the global $\alpha_s$ average  noticeably, due to an error 
  that has been largely underestimated \cite{qwg}. Using the latest development 
  in the calculation of relativistic corrections and in the treatment 
  of perturbative series in $\alpha_s$ it will be possible to obtain 
  a more appropriate determination of $\alpha_s$ from  heavy quarkonium. 
\par\noindent
{\bf Gluelump spectrum. }
In pNRQCD \cite{pnrqcd2,Bali:2003jq}  the full structure of the gluelump spectrum has been studied,
obtaining model independent predictions on the shape, the pattern, the degeneracy 
and the multiplet structure of the hybrid static energies for small $Q\bar{Q}$ 
distances that well match and interpret the existing lattice data.  
\par\noindent
{\bf Properties of baryons made of two or three heavy quarks.}
Recently the SELEX experiment has detected first signals from three-body bound states made 
 of two heavy quarks and a light one. These systems are theoretically quite interesting 
due to the interplay of HQET and NRQCD in the construction of a suitable EFT. 
 Triggered by these experimental data in \cite{3qeft}  
 EFT Lagrangians describing $QQQ$ states and $QQq$ states  have been constructed 
 and some model independent predictions on the spectra (hyperfine separations)
have been obtained.

\begin{table}[th]
\addtolength{\arraycolsep}{0.2cm}
\begin{tabular}{|c|c|c|}
\hline
~~~Ref.~~~  & ~~~~~~order~~~~~~ &  ~~~~~~~~~~~~ $m_b^{\overline{\rm MS}}(m_b^{\overline{\rm MS}})$ (GeV) ~~~~~~~~~~~~ 
\\
\hline
\cite{Beneke:1999fe} & NNLO& $4.24 \pm 0.09$ 
\\ 
\cite{Hoang:1998uv} & NNLO& $4.21 \pm 0.09$
\\
\cite{Pineda:2001zq} & NNLO & $4.210 \pm 0.090 \pm 0.025$
\\
\cite{Brambilla:2001qk}  & NNLO & $4.190 \pm 0.020 \pm 0.025$
\\
\cite{Penin:2002zv} & NNNLO &$4.349 \pm 0.070$
\\
\cite{Lee:2003hh} & NNNLO & $4.20 \pm 0.04$
\\
\cite{Contreras:2003zb} & NNNLO & $4.241 \pm 0.070$ 
\\
\hline
\hline
Ref.  & order &  $m_c^{\overline{\rm MS}}(m_c^{\overline{\rm MS}})$ (GeV)  
\\
\hline
\cite{Brambilla:2001fw}  & NNLO & $1.24 \pm 0.020$ \\
\hline
\end{tabular}
\caption{\label{tableMSmasses} Collection of recently obtained values of 
$m_b^{\overline{\rm MS}}(m_b^{\overline{\rm MS}})$ and  $m_c^{\overline{\rm MS}}(m_c^{\overline{\rm MS}})$
from the $\Upsilon$(1S) and $J/\Psi$ masses.}
\end{table}

\begin{table}[ht]
\begin{tabular}{|c|c|cccc|}
\hline
\multicolumn{6}{|c|}{$B_c$ mass ~(MeV)}\\
\hline
State &  \cite{bcexp} ~(expt) &~ \cite{Allison:2004be} ~(lattice)~ & ~ \cite{Brambilla:2000db} ~(NNLO)
~& ~\cite{Brambilla:2001fw} ~(NNLO)~& ~\cite{Brambilla:2001qk} ~(NNLO)~\\
\hline
$1^1S_0$ & $6287\pm 4.8 \pm 1.1 $ &$6304\pm12^{+12}_{-0}$ & 6326(29) & 6324(22) & 6307(17) 
\\
\hline
\end{tabular}
\caption{Different perturbative determinations of the $B_c$ mass compared with the experimental 
value and a recent lattice determination.}
\label{TabBc}
\end{table}

\vspace{-10mm}

\subsection{Strongly coupled pNRQCD}
In this case the 
matching to pNRQCD is nonperturbative \cite{pnrqcdnonpert}.
In the situation where  the other degrees of freedom 
(like those associated 
with heavy-light meson pair  threshold production and heavy hybrids) 
develop a mass gap of order $\lQ$, 
the quarkonium singlet field $\rm S$ remains as the only low energy dynamical 
degree of freedom in the pNRQCD Lagrangian (if no ultrasoft pions are considered), 
which reads \cite{pnrqcdnonpert,sw,pnrqcd2,reveft}:
\begin{equation}
\quad  {L}_{\rm pNRQCD}= {\rm Tr} \,\Big\{ {\rm S}^\dagger
   \left(i\partial_0-{{\bf p}^2 \over
 2m}-V_S(r)\right){\rm S}  \Big \} .
\end{equation}
In this regime   we recover the quark potential singlet model from
 pNRQCD.  The matching potential $V_S$ (static and relativistic corrections)
  is nonperturbative: the real part controls
the spectrum and the imaginary part controls the inclusive decays. The potential is  calculated 
in the nonperturbative matching procedure between NRQCD and pNRQCD \cite{pnrqcdnonpert,reveft}.
The great advantages of this approach include:
factorization of  hard (in the NRQCD matching coefficients) and  soft  scales
 (contained in Wilson loops or nonlocal gluon correlators);
the fact that the  low energy objects are only  glue dependent:  
this opens a window to  confinement investigations,
        on the lattice \cite{latticeval} or in QCD vacuum models \cite{rev};
the  existence of a clear  power counting indicating leading and subleading terms 
in quantum-mechanical 
perturbation theory;
the fact that the quantum mechanical  divergences (like the ones coming from iterations
    of spin delta potentials)  are absorbed  by NRQCD matching coefficients;
the definitive disappearance of  
 fake problems like the ``Lorentz structure of the potentials'';
the fact that one no longer needs
to repeat a lattice evaluation for each quarkonium state, but gets in one step the full potential 
 (which, in turn, inserted inside the Schr\"odinger equation, will 
produce all the quarkonium masses).
The calculations involve only   QCD parameters  (at some scale and in some 
scheme). 

\leftline{\bf Applications of strongly coupled  pNRQCD}
\noindent
{\bf Nonperturbative potentials and Spectrum.}
 The final result for the potential (static and
relativistic corrections) appears factorized in a part  containing the
high energy  dynamics (and calculable in perturbation theory) which is
 inherited from the NRQCD matching coefficients, and a part containing the low energy dynamics
given in terms of Wilson loops and chromo-electric and chromo-magnetic
 insertions in the Wilson loop \cite{pnrqcdnonpert}.  
The expression obtained for the potential {\it is} the QCD expression, in particular 
all the perturbative contributions to the potential 
at the hard scale are correctly taken into account.
This solves the problem of consistency with perturbative one-loop calculations
 that was previously encountered in the Wilson loop approach. Moreover, 
further contributions, including a $1/m$ nonperturbative potential,
 appear with respect to the 
Wilson loop original results \cite{rev,before3q}. The full expressions for the potentials
is given in \cite{pnrqcdnonpert}. Comprehensive phenomenological applications of these
full results are still missing.
\par\noindent
{\bf Decays.}
The inclusive quarkonium decay widths  
in pNRQCD can be  factorized  with respect to the wave function 
(or its derivatives) calculated in zero,
which  is suggestive of the early potential models results:
$
  \Gamma ({\rm H}\to{\rm LH}) = F(\als,\lQ) \, \cdot \, \vert \psi (0)\vert^2 .
$
Similar expressions hold for the electromagnetic decays.
However, 
the coefficient $F$ depends here  both on $\als$ and $\lQ$. In particular 
all NRQCD matrix elements, including the octet
ones, can be expressed through pNRQCD as products of  universal 
nonperturbative factors by the squares of the quarkonium wave functions
(or derivatives of it) at the origin. The nonperturbative factors are typically
integral of nonlocal electric or magnetic correlators and thus  
depending on the glue but not on the quarkonium state \cite{sw}. 
Typically $F$ contains both the NRQCD matching coefficients  at the 
hard scale $m$ and the nonperturbative correlators at the low energy scale $\lQ$.
The  nonperturbative correlators, being state independent, are in a
smaller number than the   nonperturbative NRQCD   matrix elements
and thus the predictive power is greatly increased in going from NRQCD to pNRQCD.
   In \cite{sw}  the inclusive decay widths into light hadrons, 
photons and lepton pairs of all $S$-wave and $P$-wave states (under threshold) 
have been calculated up to ${O}(mv^3\times (\Lambda_{\rm QCD}^2/m^2,v^2))$ 
and ${O}(mv^5)$. A  large reduction in
the number of unknown nonperturbative parameters has been  achieved and, therefore,
after having fixed the nonperturbative parameters on charmonium decays, 
new model-independent QCD predictions have been obtained  for the bottomonium decay
widths \cite{sw}.
\par\noindent
%
%

\vspace{-6mm}
\begin{theacknowledgments}
\vspace{-3mm}
I would like to thank the Organizers of Hadron05 for the perfect organization,
 the very enjoable atmosphere and the wonderful location.
\end{theacknowledgments}


\bibliographystyle{aipproc}   


\end{document}